\documentclass[
twocolumn,
preprintnumbers,
nofootinbib,
prd,aps
]{revtex4}

\usepackage{epsf}
\usepackage{latexsym}
\usepackage{amssymb}
\usepackage{epsf}
\usepackage{amsmath}
\usepackage{slashed}

\begin{document}


\newcommand{\rem}[1]{$\spadesuit${\bf #1}$\spadesuit$}

\renewcommand{\topfraction}{0.8}

\preprint{UT-HET 103}

\title{
Discriminative phenomenological features  
of scale invariant models \\for electroweak symmetry breaking 
}

\author{
Katsuya Hashino$^{(a)}$, Shinya Kanemura$^{(a)}$, Yuta Orikasa$^{(b)}$
}

\affiliation{
$^{(a)}$Department of Physics, University of Toyama,
Toyama 930-8555, Japan
\\
$^{(b)}$ School of Physics, KIAS, Seoul 130-722, Korea,\\
$^{(b)}$ Department of Physics and Astronomy, Seoul National University, Seoul
151-742, Korea
}


\begin{abstract}

Classical scale invariance (CSI) may be one of the solutions for the hierarchy problem.
Realistic models for electroweak symmetry breaking based on CSI require 
extended scalar sectors without mass terms, and the 
electroweak symmetry is broken dynamically at the quantum level by the Coleman-Weinberg mechanism. 
We discuss discriminative features of these models. 
First, using the experimental value of the mass of the discovered Higgs boson $h(125)$,   
we obtain an upper bound  on the mass of the lightest additional scalar boson ($\simeq 543$ GeV),   
which does not depend on its isospin and hypercharge.  
Second, a discriminative prediction on the Higgs-photon-photon coupling is given as a function of 
the number of charged scalar bosons, by which we can narrow down possible models   
using current and future data for the di-photon decay of $h(125)$. 
Finally,  for the triple Higgs boson coupling 
a large deviation ($\sim +70$ \%) from the SM prediction 
is universally predicted, which is independent of masses, 
quantum numbers and even the number of additional scalars. 
These models based on CSI can be well tested at LHC Run II and at future lepton colliders.  
 
\end{abstract}

\maketitle

\renewcommand{\thefootnote}{\#\arabic{footnote}}

 
By the discovery of the Higgs boson at LHC, the idea of spontaneous breaking of 
the electroweak (EW) symmetry was confirmed its correctness~\cite{HiggsDiscovery}. 
Detailed measurements of the property of the discovered Higgs particle $h(125)$ with the mass 125~GeV 
showed that the standard model (SM) with the one Higgs doublet field is   
a good description of the physics around the scale of 100 GeV 
within the uncertainty of the data~\cite{Agashe:2014kda}. 
Nevertheless, the essence of the Higgs boson and the structure of the Higgs sector remain unknown.  
The discovery of $h(125)$ provided us a step to explore the physics 
behind the EW symmetry breaking (EWSB). 
New physics beyond the SM is also required to explain phenomena such as 
dark matter, neutrino oscillation, baryon asymmetry of the universe and cosmic inflation.  
Physics of the Higgs sector is an important window to approach these problems. 

In the SM, it is known that quadratic ultraviolet divergences appear in radiative corrections 
to the Higgs boson mass, which cause the hierarchy problem~\cite{Veltman:1976rt}.  
In order to solve the problem, several new physics paradigms have been proposed 
such as supersymmetry and scenarios of dynamical symmetry breaking. 
These paradigms have been thoroughly tested by experiments. 
Simple models of dynamical symmetry breaking like Technicolor models 
have been strongly constrained by EW precision data at LEP/SLC experiments~\cite{Agashe:2014kda}. 
Supersymmetric extensions of the SM are also now being in trouble due to 
the non-observation of supersymmetric partner particles at LHC, although 
there is still hope that they can be discovered at the LHC Run II experiment.    

There is another idea that would avoid the hierarchy problem, 
which is based on the notion of classical scale invariance (CSI), 
originally proposed by Bardeen~\cite{Bardeen:1995kv}. 
In a class of models based on CSI, parameters with mass dimensions are not introduced to the Lagrangian.  
EWSB can dynamically occur via the mechanism by Coleman and Weinberg (CW)~\cite{Coleman:1973jx}, 
and masses of particles are generated by the dimensional transmutation. 
After the discovery of $h(125)$, models along this line have become popular as a possible alternative paradigm. 
The minimal scale-invariant model with one Higgs doublet has already been excluded 
by the data, so that extended scalar sectors have to be considered as realistic 
models~\cite{Gildener:1976ih,Funakubo:1993jg,Takenaga:1993ux,Lee:2012jn, 
Fuyuto:2015vna,Ishiwata:2011aa,Guo:2014bha,Endo:2015ifa}. 
In Ref.~\cite{Funakubo:1993jg,Fuyuto:2015vna},  
strongly first order EW phase transition was studied in extended Higgs models with CSI 
for a successful scenario of EW baryogenesis. 
An additional scalar field in models with CSI could be a dark matter 
if it is stable~\cite{Ishiwata:2011aa,Guo:2014bha}.   
In Ref.~\cite{Endo:2015nba}, in the CSI model with $N$-singlet fields under the $O(N)$ symmetry,  
the upper bound on $N$ was obtained from the direct search results of dark matter.   
In Ref.~\cite{Khoze:2013uia,Kannike:2015apa}, dark matter and inflation were investigated in models with CSI.   
Scale invariant models for neutrino masses were discussed in Ref.~\cite{Foot:2007ay}. 

In this Letter, we discuss discriminative phenomenological features of models for EWSB based on CSI. 
We study a full set of the models with additional scalar fields to $h(125)$, 
which can contain arbitrary number of isospin singlet fields, additional doublets or 
higher representation fields with arbitrary hypercharges. 
An unbroken symmetry may be added so that additional scalars do not have 
vacuum expectation values (VEVs).   
Otherwise, additional doublets or higher multiplets can share the VEV $v$ ($\simeq 246$ GeV) of EWSB 
 with the SM-like Higgs field.  
However, we here do not discuss the case where singlets have VEVs which 
are irrelevant to the Fermi constant $G_F$ $( \simeq 1/\sqrt{2} v^2)$~\cite{Meissner:2006zh,Ishiwata:2011aa,Khoze:2013uia}.   

First of all, a general upper bound $C$ on the mass $m_1^{\rm CSI}$ 
of the lightest scalar boson other than $h(125)$
is obtained in all models of this category, 
\begin{align}
m_1^{\rm CSI} \leq C \simeq 543 {\rm GeV}. \label{eq:upper1}
\end{align}
If we specify the structure of the model, a stronger upper bound is obtained. 
Second, a discriminative prediction on the di-photon coupling of $h(125)$ is obtained.  
In terms of the scaling factor $\kappa_{\gamma}^{\rm CSI}$ 
of the $h\gamma\gamma$ coupling, it is approximately given by 
\begin{align}
     \kappa_{\gamma}^{\rm CSI}  \simeq 1 - \frac{n}{16} - \frac{m}{4} ,  \label{eq:hgg}
\end{align}
where $n$ and $m$ are the numbers of singly- and doubly-charged scalar bosons, respectively. 
Finally, the triple Higgs boson coupling $\Gamma_{hhh}^{\rm CSI}$ 
is universally predicted at the leading order as 
\begin{align}
   \Gamma_{hhh}^{\rm CSI}  = \frac{5m_h^2}{v}  =  \frac{5}{3} \times \Gamma_{hhh}^{\rm SM tree}. \label{eq:hhh}
\end{align}
Although these results have partially been obtained in some specific models of 
CSI~\cite{Lee:2012jn,Fuyuto:2015vna,Endo:2015ifa}, 
we would like to emphasize that they are common in all models for EWSB based on CSI. 
In the following, we discuss these results in more details. 

The vacuum in these models is analyzed using the well-known method 
by Gildener and Weinberg~\cite{Gildener:1976ih}. The vacuum is surveyed along 
the flat direction, and the minimum of the effective potential can be found  
at the one-loop level by the CW mechanism~\cite{Coleman:1973jx}. 
In terms of the order parameter $\varphi$ along the flat direction, we can in general write the 
effective potential as~\cite{Gildener:1976ih}
\begin{align}
  V_{\rm eff}(\varphi) 
=  A \varphi^4 + B \varphi^4 \ln \frac{\varphi^2}{Q^2},  \label{eq:effpot}
\end{align}
where $Q$ is the scale of renormalization, and 
\begin{widetext}
\begin{align}
 A &= \frac{1}{64\pi^2 v^4} \left[ 
 3 {\rm Tr} \left( M_V^4 \ln \frac{M_V^2}{v^2}  \right)
 - 4 {\rm Tr} \left( M_f^4 \ln \frac{M_f^2}{v^2}  \right)
+  {\rm Tr} \left( M_S^4 \ln \frac{M_S^2}{v^2} \right)  
 \right],  \label{eq:coef_a}\\
B &= \frac{1}{64\pi^2 v^4} \left[ 
 3 {\rm Tr} \left( M_V^4 \right)
 - 4 {\rm Tr} \left( M_f^4 \right)
+  {\rm Tr} \left( M_S^4 \right)  
 \right], \label{eq:coef_b}
\end{align} 
\end{widetext}
where the first, the second and the third terms in the right hand side of Eqs.~(\ref{eq:coef_a}) and (\ref{eq:coef_b})
are respectively loop effects of the vector bosons, 
those of fermions, and those of extra scalar bosons~\cite{Gildener:1976ih}.  
Loop effects of $h(125)$ are not included, as they are of higher order contributions. 
Notice that we can approximately identify the SM-like Higgs 
boson $h(125)$ as the ``scalon"~\cite{Gildener:1976ih}.   
From the stationary condition, 
\begin{align}
\left. \frac{\partial V_{\rm eff}}{\partial \varphi} \right|_{\varphi=v} =0 ,
\end{align}
we obtain 
\begin{align}
\ln \frac{v^2}{Q^2} = - \frac{1}{2} - \frac{A}{B},   \label{eq:logq}
\end{align}
by using which, the mass of $h(125)$ is obtained as 
\begin{align}
m_h^2 \equiv \left. \frac{\partial^2 V_{\rm eff}}{\partial \varphi^2}\right|_{\varphi=v} = 8Bv^2\;
 \simeq (125 {\rm GeV})^2. 
\label{eq:mass}
\end{align}

In the case where we only extend the scalar sector and do not extend the vector boson sector nor the 
fermion sector, Eq.~(\ref{eq:mass}) can be rewritten as\footnote{Existence of chiral fourth generation 
fermions have already been excluded by experiments~\cite{Kribs:2007nz,Hashimoto:2010at}. 
Hence, we do not consider additional fermions.
}  
\begin{align}
{\rm Tr} M_S^4  = 8\pi^2 v^2 m_h^2 -3m_Z^4 -6m_W^4 + 12m_t^4 \; ( \equiv C^4).
\end{align} 
Because all quantities in the right hand side are known from the current data~\cite{Agashe:2014kda}, 
this equation gives the constraint on the scalar sector. 
When the scalar sector contains $N$ scalar bosons in addition to $h(125)$,  
masses of these extra bosons can be written as 
$m_1 \leq m_2 \leq \cdots \leq m_N$, where $m_i$ is the mass of the $i$-th scalar boson. 
We then obtain an upper bound on the mass $m_1^{\rm CSI}$ of the lightest scalar boson other than 
$h(125)$ as 
\begin{align}
   m_1^{\rm CSI} 
    &\leq \frac{C}{\sqrt[4]{N_{0,0} + 2 N_{0,1} + 4 N_{\frac{1}{2}, \frac{1}{2}} + 3 N_{1, 0} + 6 N_{1,1} + \cdots }}, 
\end{align}
where $N_{I,Y}$ is the number of additional scalar fields with isospin $I$ and hypercharge $Y$. 
Since  $N_{0,0} + 2 N_{0,1} + 4 N_{\frac{1}{2}, \frac{1}{2}} + 3 N_{1, 0} + 6 N_{1,1} + \cdots \geq 1$,  we obtain 
the general upper bound $C (\simeq 543$ GeV) as given in Eq.~(\ref{eq:upper1}).
This bound is given for all models for EWSB based on CSI  with extended scalar bosons.  

If we specify models, for example,  to those with only doublets, we obtain the stronger bound as \begin{align}
 m_1^{\rm CSI} \leq \frac{C}{\sqrt[4]{4 N_{\frac{1}{2}, \frac{1}{2}}} }  \sim  \frac{1}{\sqrt[4]{N_{\frac{1}{2},\frac{1}{2}}}} 
 \times  383 {\rm GeV} .\label{eq:upper2}
\end{align} 
In Ref.~\cite{Lee:2012jn}, the similar bound  has been discussed for the case of $N_{\frac{1}{2},\frac{1}{2}} =1$ 
in addition to the unitarity bound.


Next, the decoupling theorem~\cite{Appelquist:1974tg} states that 
quantum effects of heavy particles on low-energy observables 
decouple in the large mass limit. 
However, this is not the case for the models for EWSB based 
on CSI, where all massive particles 
obtain their masses from $v$, the VEV of EWSB.  In such a case, 
significant non-decoupling effects can 
cause large deviations in low energy observables from their SM predictions. 
 
 As an example, let us discuss the one-loop induced 
 coupling $h\gamma\gamma$ in models 
 with $n$ singly-charged scalar bosons and $m$ doubly-charged ones. 
 The ratio of the decay rate $\Gamma_{h\to\gamma\gamma}^{(n,m)}$ to the SM value
 $\Gamma_{h\to\gamma\gamma}^{\rm SM}$ 
 is given at the one-loop level by using the well-known formula in Ref.~\cite{Shifman:1979eb} as  
 \begin{widetext}
\begin{align}
\frac{\Gamma^{(n,m)}_{h\to\gamma\gamma}}{\Gamma^{\rm SM}_{h\to\gamma\gamma}}
 \sim \left|1 + \frac{1}{2} \frac{ \sum_{i=1}^n (v/m_{\phi_i^\pm}^2) \lambda_{h \phi_i^+ \phi_i^-} A_0(\tau_{\phi_i^\pm}) 
 + 4 \sum_{j=1}^m (v/m_{\phi_j^{\pm\pm}}^2) \lambda_{h \phi_j^{++} \phi_j^{--}} A_0(\tau_{\phi_j^{\pm\pm}}) }
 {A_1(\tau_W^{}) +\frac{4}{3} A_{\frac{1}{2}} (\tau_t)}
  \right|^2 , \label{eq:shifman}
\end{align}
\end{widetext}
where $\tau_x = 4m_x^2/m_h^2$ with $m_x$ being the mass of $x$,  and 
\begin{align}
  A_1(\tau_W^{}) &= - 2\tau_W^2\left\{2\tau_W^{-2} + 3 \tau_W^{-1} + 3 (2 \tau_W^{-1} -1) f(\tau_W^{-1})\right\},\\
  A_{\frac{1}{2}}(\tau_t)   &= 2 \tau_t^2 \left\{ \tau_t^{-1} +(\tau_t^{-1} -1) f(\tau_t^{-1})\right\},\\
  A_0(\tau_i) &= -\tau_i^2 \left\{\tau_i^{-1} -f(\tau_i^{-1})\right\}
\end{align}
with
\begin{align}
  f(z) = \left\{  
  \begin{array}{ll} 
    \arcsin^2\sqrt{z} & z\leq 1 \\
    -\frac{1}{4} 
     \left( \ln\frac{1+\sqrt{1-1/z}}{1-\sqrt{1-1/z}} - i \pi \right)    & z >1 \\ 
    \end{array}
    \right. .
\end{align}

\begin{figure}[t]
  \centerline{\epsfxsize=0.475\textwidth\epsfbox{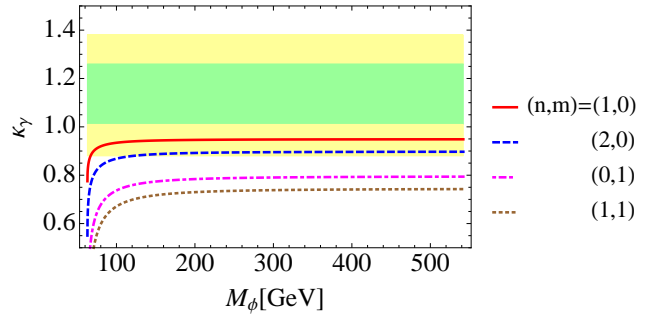}}
  \caption{The prediction on $\kappa_{\gamma}$ in the models for EWSB 
  based on CSI. The models with $(n,m)=(1,0)$, $(2,0)$, $(0,1)$ and $(1,1)$ are 
  shown, where $n$ ($m$) is the number of singly- (doubly-) charged scalar bosons. 
  The regions allowed by the current data from CMS~\cite{k_gamma_cms} 
  are also indicated at the 1 $\sigma$ and 2$\sigma$ levels. }
  \label{fig}
\end{figure}

In models for EWSB based on CSI, the coupling constants of  
$h\phi_i^+\phi_i^-$ and $h\phi_i^{++}\phi_i^{--}$  are given by  
\begin{align}
\lambda_{h \phi_i^+ \phi_i^-} = \frac{2 m_{\phi_i^\pm}^2}{v},  \;\;
\lambda_{h \phi_i^{++} \phi_i^{--}} =  \frac{2 m_{\phi_i^{\pm\pm}}^2}{v},  \label{eq:nondec}
\end{align} 
From Eqs.~(\ref{eq:shifman}) and (\ref{eq:nondec}), 
the scaling factor $\kappa_\gamma^{\rm CSI}$ is calculated as approximately given 
in Eq.~(\ref{eq:hgg}), which is valid in the large mass limit.  
By comparing this result with the current data from LHC,   
$\kappa_{\gamma} = 1.14^{+0.12}_{-0.13}$ (CMS)~\cite{k_gamma_cms} 
and $1.19^{+0.15}_{-0.12}$ (ATLAS)~\cite{k_gamma_atras}, 
the number of charged scalar bosons is strongly constrained.
In Fig.~\ref{fig}, we show predicted values of $\kappa_\gamma$ in models for EWSB based on CSI 
as a function of the common mass of charged scalar bosons. 
We learn that models with at most only one or two kinds of singly-charged scalar 
bosons are allowed. Scale invariant models for EWSB 
with doubly charged scalar bosons have been already excluded.

At the LHC Run II experiment, the data for Higgs coupling measurements will be 
drastically improved, where $\kappa_\gamma$ 
can be measured with the  $5$-$7$ \% accuracy~\cite{Dawson:2013bba}, so that      
the number of charged particles can be determined in models for EWSB based on CSI.  
If existence of charged scalar bosons is excluded, only the models with neutral singlets are allowed,  
while if the existence of one or two singly-charged scalar boson is indicated from the future 
di-photon decay data, we can complementarily test it by direct searches for charged scalar bosons. 
In any case, we can largely constrain the models for EWSB based on CSI at the 
LHC Run II experiment.

For the result on the $hhh$ coupling in Eq.~(\ref{eq:hhh}), we start 
from discussing the case of the the SM. 
The quantum effect of the top quark  is given approximately in~\cite{Kanemura:2004mg} 
\begin{align}
  \Gamma^{\rm SM}_{hhh} \simeq \frac{3 m_h^2}{v} \left( 1 - \frac{m_t^4}{\pi^2 v^2 m_h^2} + \cdots \right), 
\end{align}
where the quartic power-like top mass contribution $m_t^4$ appears. 
These loop effects remain after the renormalization of the mass parameter 
($\mu^2$) and the quartic coupling constant ($\lambda$) in the lowest-order effective potential 
\begin{align}
V(\varphi)_{\rm tree}  = - \frac{1}{2}\mu^2 \varphi^2 + \frac{1}{4} \lambda \varphi^4. 
\end{align} 
We note that similar non-decoupling effects can also appear for bosonic loop contributions 
in some cases in massive extended Higgs models~\cite{Kanemura:2004mg}.

On the other hand, in the models for EWSB based on CSI 
where no mass term is introduced, the situation is drastically changed. 
From the effective potential in Eq.~(\ref{eq:effpot})  with the relations in 
Eqs.~(\ref{eq:logq}) and (\ref{eq:mass}), we obtain 
\begin{align}
\Gamma_{hhh}^{\rm CSI} \equiv \left. \frac{\partial^3 V_{\rm eff}}{\partial \varphi^3}\right|_{\varphi=v}  = 40 v B = \frac{5m_h^2}{v}. 
\end{align}
Using the tree-level SM coupling 
$\Gamma_{hhh}^{\rm SM tree}$ ($= 3m_h^2/v$), we obtain the result in Eq.~(\ref{eq:hhh}). 
Because of CSI, the way of renormalization is different  
from the case with massive theories. 
Consequently,  
the renormalized $hhh$ coupling is expressed only in terms of $m_h$ and $v$. 
The deviation from $\Gamma_{hhh}^{\rm SM tree}$ is about +67\%. 
This is universal for all models for EWSB based on CSI.  
Notice that this universality  is broken 
in the higher order calculation, depending on details of the scalar sector of each model, 
although the difference is not so large, as discussed in Ref.~\cite{Endo:2015ifa} 
for a specific CSI model with O($N$) singlets.  
Therefore, we can test the models for EWSB based on CSI by using 
the $hhh$ coupling, which can be measured  
with the 13\% accuracy 
at the  International Linear Collider (ILC)~\cite{Asner:2013psa}.


Finally, some comments on the phenomenological consequences are in order. 
The upper bound ($\simeq 543$ GeV) on the mass of the lightest scalar boson 
other than $h(125)$  holds generally whatever its representation and charge. 
Thus, we might expect that the second scalar boson in 
scale invariant models for EWSB can be discovered at current and future LHC experiments. 
However, the detectability strongly depends on the detail of each model.
By future measurements of $\kappa_\gamma$ at LHC Run II, the number of singly-charged 
scalar bosons can be determined. 
If at least one charged scalar boson is indicated, 
the possibility of a scalar sector with extra doublets would be high. 
For the case with a multi-doublet structure, the upper bound on $m_1^{\rm CSI}$ is 
stronger, as shown in Eq.~(\ref{eq:upper2}). 
The detectability of additional scalars is very high especially when they couple to 
quarks and leptons as studied by many authors~\cite{LHCdetect}.   
Even if they do not have Yukawa interaction due to their inert property, 
we may detect them  
at LHC Run II~\cite{Dolle:2009ft} 
or at lepton colliders like the ILC~\cite{Aoki:2013lhm}. 
We can then finally discriminate the models from usual (massive) multi-Higgs doublet models  
by measuring the $hhh$ coupling at the ILC and by testing the prediction in Eq.~(\ref{eq:hhh}). 
On the other hand, if future data for $\kappa_\gamma$ indicate that there is no charged scalar boson, 
the scalar sector is composed of only singlets as additional scalar fields. 
The testability at LHC Run II is then unclear even if their masses are light enough. 
Still, we can definitely test the models by measuring the $hhh$ coupling at the ILC. 
A detailed study is performed elsewhere~\cite{hko2}.

We have discussed general aspects of models for EWSB with CSI. 
There is a general upper bound on the mass of the lightest scalar boson other than $h(125)$.  
The deviation in the $h\gamma\gamma$ coupling is mostly determined by the number of charged scalars. 
The deviation in the $hhh$ coupling from the SM prediction is universally about $ +70$ \% in these models.
By using these results, the set of models based on CSI can be well tested 
at LHC and the ILC.  

{\it Acknowledgment}: 
We would like to thank Michio Hashimoto,  
Mariko Kikuchi and Hiroaki Sugiyama for useful discussions. 
This work was supported, in part,  by Grant-in-Aid for Scientific Research No.\ 23104006 (SK), 
Grant H2020-MSCA-RISE-2014~no.~645722 (Non Minimal Higgs) (SK), 
and 
NRF Research No. 2009-0083526 of the Republic of Korea  (YO).


\end{document}